\documentclass[10pt,showkeys,aps,prl,twocolumn,floatfix,superscriptaddress]{revtex4-1}

\usepackage{graphicx} 
\usepackage{hyperref}  
\usepackage{amssymb}  
\usepackage{amsmath} 
\usepackage{array}
\usepackage{multirow}
\usepackage{color}

\begin{document}

%\begin{frontmatter}

\title{Search for a bound di-neutron by comparing \texorpdfstring{$^3$He}{3He}(e,e'p)d and \texorpdfstring{$^3$H}{3H}(e,e’p)X measurements}

%
% Key Student and Postdoc (checking with Eli on the order)
%
\author{D.~Nguyen} \thanks{Equal Contribution}
\affiliation{Jefferson Lab, Newport News, Virginia 23606, USA}
\affiliation{Massachusetts Institute of Technology, Cambridge, MA 02139, USA}

\author{C.~Neuburger} \thanks{Equal Contribution}
\affiliation{School of Physics and Astronomy, Tel Aviv University, Tel Aviv 6997845, Israel}

\author{R.~Cruz-Torres}
\affiliation{Lawrence Berkeley National Laboratory, Berkeley, CA 94720}

\author{A.~Schmidt}
\affiliation{Massachusetts Institute of Technology, Cambridge, MA 02139, USA}
\affiliation{George Washington University, Washington, DC, 20052}

\author{D.W.~Higinbotham}
\affiliation{Jefferson Lab, Newport News, Virginia 23606, USA}

\author{J.~Kahlbow}
\affiliation{Massachusetts Institute of Technology, Cambridge, MA 02139, USA}
\affiliation{School of Physics and Astronomy, Tel Aviv University, Tel Aviv 6997845, Israel}

\author{P.~Monaghan}
\affiliation{Christopher Newport University, Newport News, VA, 23606, USA}

\author{E.~Piasetzky}
\affiliation{School of Physics and Astronomy, Tel Aviv University, Tel Aviv 6997845, Israel}

\author{O.~Hen}
\affiliation{Massachusetts Institute of Technology, Cambridge, MA 02139, USA}

\begin{abstract}
We report on a search for a bound di-neutron by comparing electron-induced proton-knockout  $(e,e'p)$ measurements from Helium-3 ($^3$He) and Tritium ($^3$H). The measurements were performed at Jefferson Lab Hall A with a 4.326 GeV electron beam, and kinematics of large momentum transfer ($\langle Q^2 \rangle \approx 1.9$ (GeV/$c$)$^2$) and $x_B>1$, to minimize contributions from non quasi-elastic (QE) reaction mechanisms. Analyzing the measured $^3$He missing mass ($M_{miss}$) and missing energy ($E_{miss}$) distributions, we can distinguish the two-body break-up reaction, in which the residual proton-neutron system remains bound as a deuteron. In the $^3$H mirror case, under the exact same kinematic conditions, we do not identify a signature for a bound di-neutron with similar binding energy to that of the deuteron. We calculate exclusion limits as a function of the di-neutron binding energy and find that, for binding equivalent to the deuteron, the two-body break-up cross section on $^3$H is less than 0.9\% of that on $^3$He in the measured kinematics at the 95\% confidence level. 
%With model dependent assumptions, 
This limit implies that the di-neutron content of the tritium spectral function is less than 1.5\%.

\end{abstract}

\maketitle
%=============Intro PRL compatible ===================%
\section{Introduction}
Neutrons ($n$) and protons ($p$) are the building blocks of atomic nuclei. 
Their lightest bound system is the deuteron, made from one proton and one neutron. 
While the deuteron is bound by 2.2~MeV, it appears that in contrast its charge-symmetric partners, the neutron-neutron and proton-proton systems, do not form bound states. 
However, calculations from first principles like in Quantum Chromodynamics~\cite{Beane:2011iw}, or pion-less Effective Field Theory~\cite{Hammer:2014rba} do not rule out a bound di-neutron system.

Searches for a fully neutral multi-neutron system such as $2n$, $3n$, or $4n$ have extensive history and have sparked large interest in both experimental and theoretical studies~\cite{Kundu:1948,Cohen:1953,Glasgow:1967zz,Spyrou:2012zz,Kisamori:2016jie}. 
If such a multi-neutron system is to be observed it will have far reaching consequences on our description of nucleon-nucleon interactions, the structure of nuclei, and possibly even Big Bang nucleosynthesis~\cite{Witala:2010ky,Hammer:2014rba,Kneller:2003ka}.

Direct scattering experiments between two neutrons are impossible to perform due to the lack of a stable neutron target.
The scattering length, $a_{nn}$, in the $^1S_0$ singlet state is large and negative, indicating no bound di-neutron system. However, various (indirect) measurements have yielded inconsistent values~\cite{GonzalezTrotter:1999zz,Huhn:2000zz}, making it almost bound. 
Early indirect searches in nuclei led to contradicting results, claiming to have found evidence for a fragile bound two-neutron system~\cite{Kundu:1948} or showing negative results~\cite{Cohen:1953,Glasgow:1967zz}.

Here we report on a study using a new technique of precision electron induced hard proton knockout from $^3$H, to access the residual two-neutron system. 
Hard proton knockout from $^3$He with a residual deuteron system is measured simultaneously and serves as a control system.

The $^3$H nucleus used here presents the ideal system as the $nn$ system might be pre-formed in its ground state in the presence of only one additional proton. 
The hard removal of the proton with large momentum transfer minimizes distortions of the $nn$ system and conserves the initial state separation between the hypothetical di-neutron. 
This is in contrast to scattering off a deuteron as had been used in many previous studies~\cite{Witala:2012te}. 
%The latter reactions need embedding of a neutron or a charge exchange of the proton to a neutron. 
The use of $^3$H target allows to study the two-neutron system without the need to measure neutrons directly. 
The knocked-out proton is leaving with high energy similarly to what has been proposed 
for inverse kinematics in Ref.~\cite{goebel:2021}.

%==============================Experimental section ==========================//
\section{Experiment}
The experiment was performed in 2018 at Hall A of the Thomas Jefferson National Accelerator Facility.
A 20 $\mu$A continuous wave electron beam with an energy of 4.326 GeV was directed alternately on one of four identical 25-cm long gas target cells that were filled with Hydrogen ($70.8\pm 0.4$ mg/cm$^2$), Deuterium ($142.2 \pm 0.8$ mg/cm$^2$), Helium-3 ($53.4 \pm 0.6$ mg/cm$^2$), and Tritium ($85.1 \pm 0.8$ mg/cm$^2$)~\cite{Santiesteban:2018qwi}. Only data collected from the Helium-3 and Tritium targets were used in this work. 

Two nearly identical high-resolution spectrometers (HRS)~\cite{Amroun:1994qj}, labeled left and right with respect to the beam direction, were used to detect quasi-elastic $(e,e'p)$ events.
Each HRS consisted of three quadrupole magnets for focusing and one dipole magnet for bending the trajectory of the particles to transport them from the interaction region to the detector package. 
The detector package in each HRS was composed of two vertical drift chambers used for tracking and two scintillation counter planes that provided timing and trigger signals. A CO$_2$ Cherenkov detector placed between the scintillators was used to separate electrons and pions, and a lead-glass calorimeter placed after them was used for further particle identification. 
This configuration is the same as in Refs.~\cite{Cruz-Torres:2020uke, Cruz-Torres:2019bqw} and slightly updated with respect to the one in Ref.~\cite{Alcorn:2004sb}.

Scattered electrons were detected in the left-HRS at central kinematic setting of momentum $\vert\vec p_e{\!'}\vert =3.543$~GeV/$c$ and angle $\theta_{e} = 20.88^\circ$ corresponding to an energy transfer $\omega = 0.78$ GeV, central four momentum transfer $Q^2 = \vec{q}^{2} - \omega^{2} = 2.0$ (GeV/$c$)$^2$ (where the transfer momentum vector is $\vec{q} = \vec{p}_{beam}- \vec{p}_e$) , and Bjorken $x_{B} = Q^2/2m_p\omega =1.4$ (where $m_{p}$ is the proton mass). Knocked-out protons were detected in the right-HRS at two different central kinematic settings, $(p_{p}, \theta_{p})$ = (1.481 GeV/$c$, 48.82$^\circ$) and (1.246 GeV/$c$, 58.50$^\circ$) referred to here as low-$p_{miss}$ and high-$p_{miss}$, respectively, where $\vec{p}_{miss} = \vec{p}_p - \vec{q}$. Only data collected from the low-$p_{miss}$ setting ($40\leq p_{miss} \leq 250$ MeV/$c$) was used in this work since it has a larger two-body-breakup contribution in $^3$He compared to the high-$p_{miss}$ setting.
%In the plane wave impulsive approximation (PWIA), the missing momentum and missing energy equal to the initial momentum and separation energy of knocked-out nucleon, $\vec{p}_i = \vec{p}_{miss}$ and $E_i = E_{miss}$.
The missing energy was defined as $E_{miss}=\omega - T_p - T_{A-1}$, where $T_{A-1} = (\omega + m_A - E_p) - \sqrt{(\omega + m_A - E_p)^2 - |\vec{p}_{miss}|^2}$ is the reconstructed kinetic energy of the residual $A-1$ system. $T_p = E_p - m_p$ and $E_p$ are the kinetic and total energy of the detected proton respectively. This expression of missing energy includes any binding energy lost in removing a proton from the target nucleus. The missing mass was defined as $M_{miss} = \sqrt{(\omega + m_A - E_p)^2 - |\vec{p}_{miss}|^2}$. In the quasi-elastic $(e,e'p)$ scattering of $^3$He, the final states can result in either two body break up (2bbu) $pd$ or three body break up (3bbu) $ppn$ corresponding to the threshold energies of $E_{miss}$ $\sim$5.5 MeV and $\sim$7.7 MeV, respectively. For $^3$H, the three-body final state, $pnn$, corresponds to $E_{miss}$ threshold of $\sim$8.5 MeV.

%==========================Analysis section =============================//
\section{Analysis}
The data analysis follows exactly the same event selection criteria as the analyses reported in the Refs.~{\cite{Cruz-Torres:2019bqw}} and \cite{Cruz-Torres:2020uke}. Electron candidates were required to deposit at least a half of their energy in the calorimeter ($E_{cal}/|\vec{p}| > 0.5$). Coincident $(e,e'p)$ events were selected by applying a $\pm3\sigma$ cut around the relative electron and proton event times. The random-coincidence event rate was negligible due to the low luminosity of this experiment.

In order to exclude scattering events from the target wall, the reconstructed electron vertex position was required to be within 9 cm from the target center (the target walls were located at $\pm 12.5$~cm). In addition, a $\pm3\sigma$ cut was applied around the relative electron and proton reconstructed target vertices (corresponding to $\pm1.2$~cm cut). The target wall contribution was determined using the empty-target measurement and proved to be negligible ($\ll1$\%) 

We only selected events that were detected within $\pm4$\% of central spectrometer momentum, and $\pm27.5$ mrad in-plane angle and $\pm55.0$ mrad out-plane angle relative to the center of spectrometer acceptance. To minimize final state interactions (FSIs), an additional restriction on the angle between the recoiling vector ($\vec{p}_{recoil} = - \vec{p}_{miss}$) and $\vec{q}$, i.e., $\theta_{rq} < 37.5^\circ$ was applied \cite{HallA:2011gjn}.

For each nuclear target, the normalized yield, $Y$, is defined as:
\begin{equation}
    Y = \frac{N}{Q \cdot f_{lt} \cdot \rho \cdot b},
\end{equation}
where $N$ is the number of events that passed all the selection cuts, $\rho$ is nominal areal density of the gas in the target cell, $Q$ is total accumulated beam charge, $f_{lt}$ is the live-time fraction in which the detectors are able to collect data and $b$ is a boiling correction factor to account for changes in the target density caused by local beam heating. The accumulated beam charge was determined by the Hall A beam current monitors, with an accuracy of better than 1\%. The live-time fraction was monitored by the data acquisition system with negligible uncertainty. The target gas density was estimated from the temperature and pressure when the cells were filled, and is the dominant source of normalization uncertainty. The boiling correction was determined by measuring the beam current dependence of the inclusive event yield \cite{Santiesteban:2018qwi}.

In this analysis, we extract the ratio of the possible di-neutron signal to the total detected 2bbu yield in $^3$He at the same kinematics using exactly the same analysis cuts. In order to perform this comparison, we need the relative normalization between the two data sets, $n$ defined by:
\begin{equation}
n=\frac{\rho_{^3{\rm H}} \cdot Q_{^3{\rm H}} \cdot f_{lt,^3{\rm H}} \cdot b_{^3{\rm H}}}{\rho_{^3{\rm He}} \cdot Q_{^3{\rm He}} \cdot f_{lt,^3{\rm He}} \cdot b_{^3{\rm He}}},
\end{equation}
We estimate $n$ to be $1.632 \pm 0.041$, i.e., an uncertainty on the relative normalization of 2.5\%. 

From the time the tritium target was filled to the time the experiment was conducted, a small fraction of the tritium nuclei had decayed into helium-3. This $^3$He contamination poses a background for the di-neutron search, since 2bbu break-up events from $^3$He cannot be distinguished from 2bbu (di-neutron) events from $^3$H. For this experiment, previous analyses~\cite{Cruz-Torres:2019bqw,Cruz-Torres:2020uke} determined the helium contamination to be 
$c=2.78\pm 0.18$\%, and we used this number to normalize the $^3$He spectrum in order to model the contamination background during $^3$H running.

\begin{figure*}[ht]
    \centering
    \includegraphics[width = 0.45\linewidth]{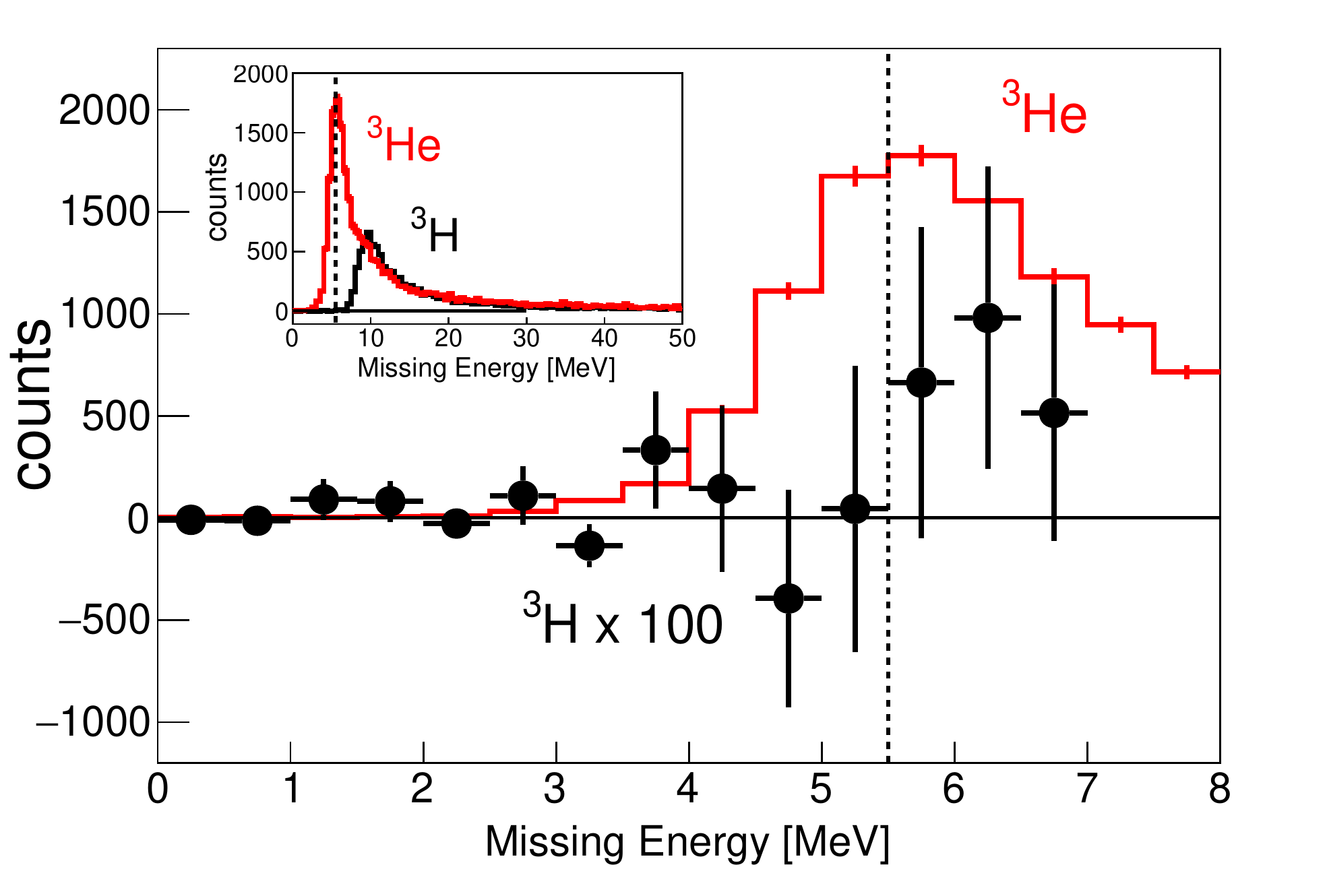}
    \includegraphics[width = 0.45\linewidth]{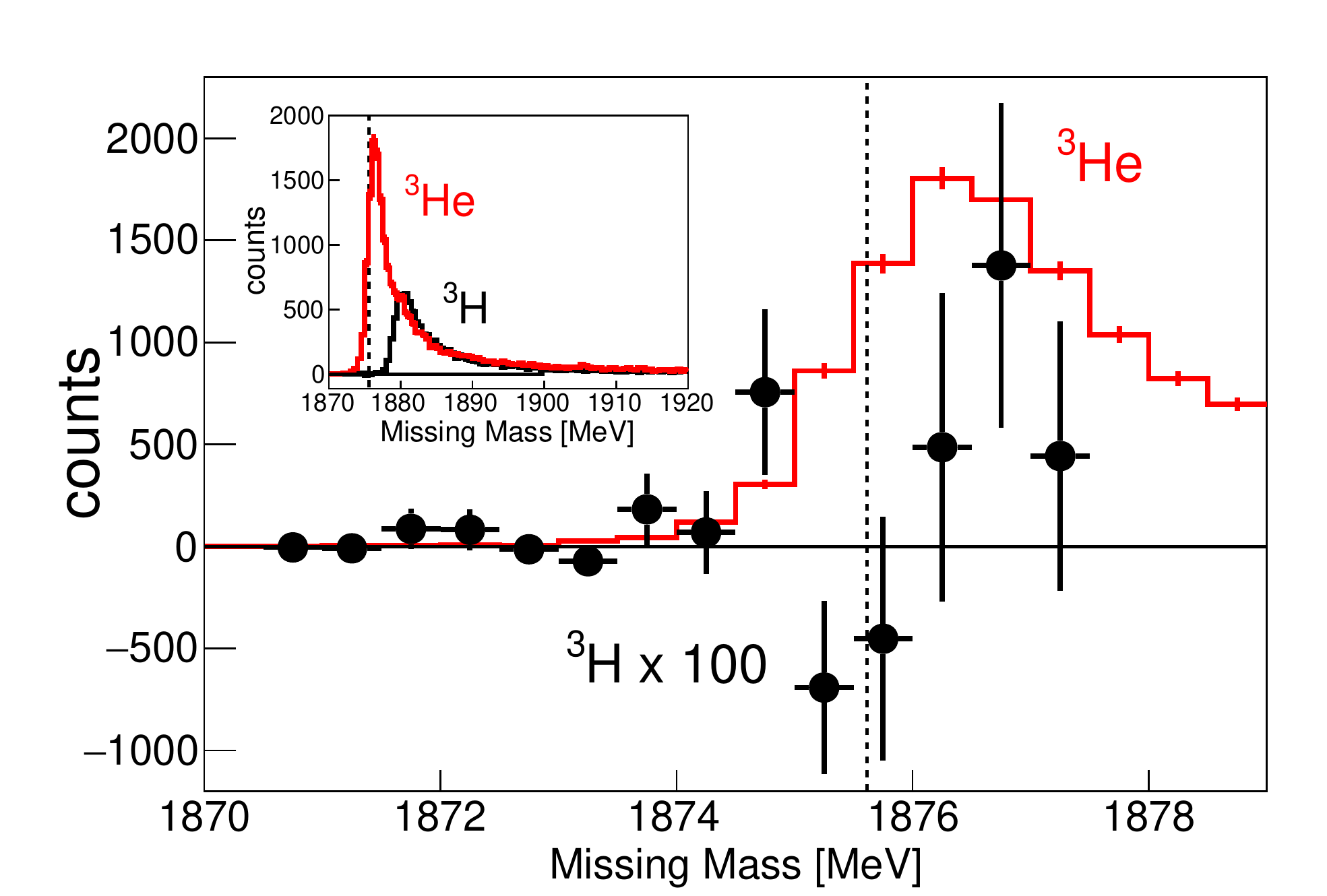}
    \caption{Left plot is the missing energy distribution of $^3$He (red) and $^3$H scaled by 100 (black dots). Inset: A larger range of the same plots without 100 times scaling on $^3$H. The dashed vertical line corresponds to the $^3$He $2bbu$ energy of 5.5 MeV. Right plots is the corresponding missing mass distribution of $^3$He (red) and $^3$H scaled by 100 (black dots). Inset: A larger range of the same plots without 100 times scaling on $^3$H. The dashed vertical line corresponds to the mass of the deuteron 1875.6 MeV.}
    \label{Emiss-result}
\end{figure*}

The tritium three-body break-up threshold occurs at a missing energy of 8.5 MeV. Based on the resolution of the spectrometers, we chose $E_{miss}<7$~MeV to be the signal region for the di-neutron search. The measured and relatively normalized missing energy and missing mass spectra for $^3$He and $^3$H after background subtraction are shown in Fig.~\ref{Emiss-result}. In the signal region, we observe 21 $^3$H counts above background with a statistical uncertainty of $\pm 16$.

In order to set limits on the cross section for producing bound di-neutron we took into account the uncertainty from systematic effects such as the contamination of $^3$He events in the $^3$H di-neutron signal region. We performed a likelihood analysis with one parameter of interest, $R$, defined as the ratio of the di-neutron cross section $\sigma_{nn}^{^3\text{H}}$ on $^3$H to the 2bbu cross section, $\sigma_{2bbu}^{^3\text{He}}$, on $^3$He. Systematic effects were considered by including four nuisance parameters: the true average count rate for 2bbu on $^3$He, $\lambda_{2bbu}^{^3\text{He}}$, the true normalization factor between the $^3$H and $^3$He data sets, $n_0$, the true helium contamination in the tritium target, $c_0$, and the true relative efficiency for detecting $^3$H and $^3$He 2bbu events, $\epsilon_0$. This last factor is necessary because the unknown binding energy of the di-neutron. For a di-neutron that is barely bound, much of the 2bbu signal will fall outside of our di-neutron signal region. We estimated this efficiency using a data-driven method: we used the shape of the $^3$He $E_{miss}$ spectrum as a template. By shifting this spectrum and studying the change in the number of counts in the 2bbu signal region, the relative efficiency for different possible di-neutron binding energies can be estimated. This relative efficiency ranges from 100\% for a di-neutron with binding energy equal to that of a deuteron to $\approx 0\%$ for a di-neutron with zero binding energy, which is indistinguishable in our measurement from 3bbu. 

\begin{figure}[ht]
    \centering
    \includegraphics[width = 0.92\linewidth]{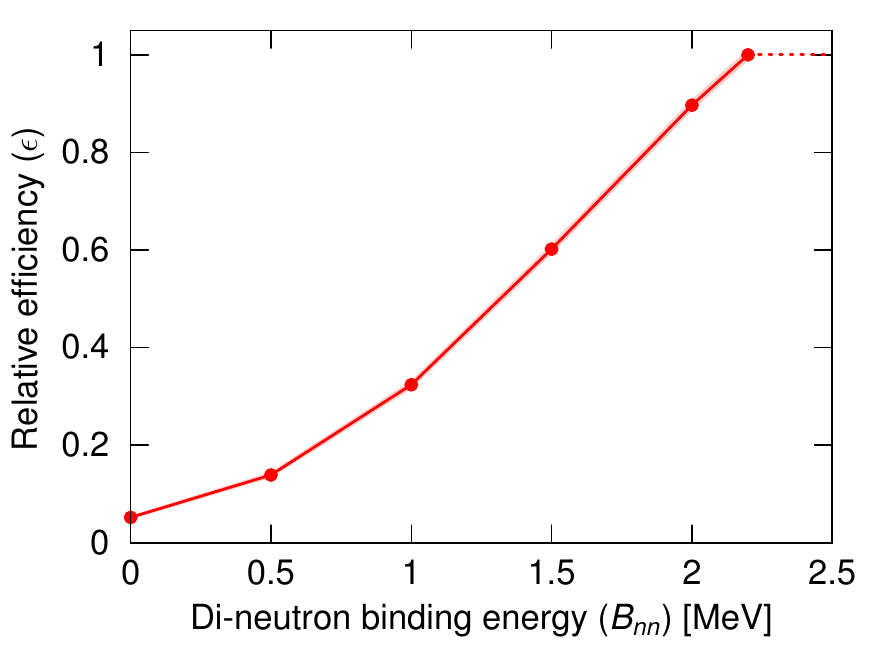}
    \caption{The relative efficiency for observing a dineutron event in $^3$H in the signal region relative to the efficiency for detecting a 2bbu event in $^3$He in an equivalent signal region. For large di-neutron binding energies, our data-driven estimate can only return a lower bound, since the method would become sensitive to leakage from 3bbu events in $^3$He. }
    \label{fig:eff}
\end{figure}

Our estimate of the relative efficiency, $\epsilon$, for a given di-neutron binding energy, $B_{nn}$, is calculated
\begin{equation}
    \epsilon = \frac{\int_0^{6.2\text{~MeV} - B_d + B_{nn}}
    N^{^3\text{He}} dE_m}
    {\int_0^{6.2\text{~MeV}} N^{^3\text{He}} dE_m},
\end{equation}
where $B_d$ is the binding energy of the deuteron, i.e., 2.2 MeV, and $B_{nn}$ is the binding energy of the possible di-neutron. By definition, the relative efficiency is 1 for a di-neutron bound by the same energy as the deuteron. The integration limit of 6.2~MeV corresponds to the 7.0~MeV di-neutron signal region, adjusted for the 0.8~MeV difference in binding energy between $^3$He and $^3$H. The estimated relative efficiency is shown in Fig.~\ref{fig:eff}. We cannot estimate the relative efficiency for detecting di-neutrons bound by more than 2.2~MeV, since there will be unavoidable contamination from 3bbu events in $^3$He. We can set a lower bound, which is indicated by the dashed line. The range studied at low binding energies is the most relevant one considering consistency with other measurement for instance the neutron-deuteron breakup cross sections~\cite{Witala:2010ky}.

Given a set of guess values for the parameters $R$, $\lambda_{2bbu}^{^3\text{He}}$, $n_0$, $c_0$, and $\epsilon_0$, the likelihood of having the measured data of $N^{^3\text{H}}_{2bbu}$ counts in the $^3$H di-neutron signal region, $N^{^3\text{He}}_{2bbu}$ counts in the corresponding 2bbu signal region in $^3$He, the measured relative normalization between the data-sets, $n$, and our data-driven relative efficiency estimate, $\epsilon$, is:
\begin{multline}
    L = P(N_{2bbu}^{^3\text{He}}|\lambda_{2bbu}^{^3\text{He}}) \cdot
    P(N_{2bbu}^{^3\text{H}}|n_0 \lambda_{2bbu}^{^3\text{He}} (c_0 + \epsilon_0 R)) \\
    \cdot G(n-n_0|\sigma_n) \cdot G(c-c_0|\sigma_c) \cdot
    G(\epsilon-\epsilon_0|\sigma_\epsilon),
\end{multline}
where $P$ represents a Poisson distribution, $c$ represents the 2.78\% contamination fraction of $^3$He in the $^3$H target cell, and $G$ represents a Gaussian distribution, where $\sigma_n$ represents the uncertainty on the relative normalization, $\sigma_c$ represents the uncertainty on the helium contamination, and $\sigma_\epsilon$ represents the uncertainty on the relative efficiency. We determined exclusion limits on $R$ based on the change in log-likelihood, $\Delta \log L$, while finding optimal values of $\lambda_{2bbu}^{^3\text{He}}$, $n_0$, $c_0$, and $\epsilon_0$ for each value of $R$. For example, to estimate the exclusion at the $2\sigma$ or 95\% confidence level, we solved for the value of $R$ at which $\Delta \log L = 2$.

\section{Results}

The missing energy and missing mass spectra measured from the $^3$H after subtracting the 2.78\% $^3$He contamination are shown in Figs.~\ref{Emiss-result}, along with the corresponding relatively normalized spectra from $^3$He. The $^3$He spectra clearly shows the 2bbu signal, in which a bound deuteron remains. The $^3$H spectra do not show a statistically significant 2bbu signal, meaning that we do not see evidence of a bound dineutron.

\begin{figure}[ht]
    \centering
    \includegraphics[width = \linewidth]{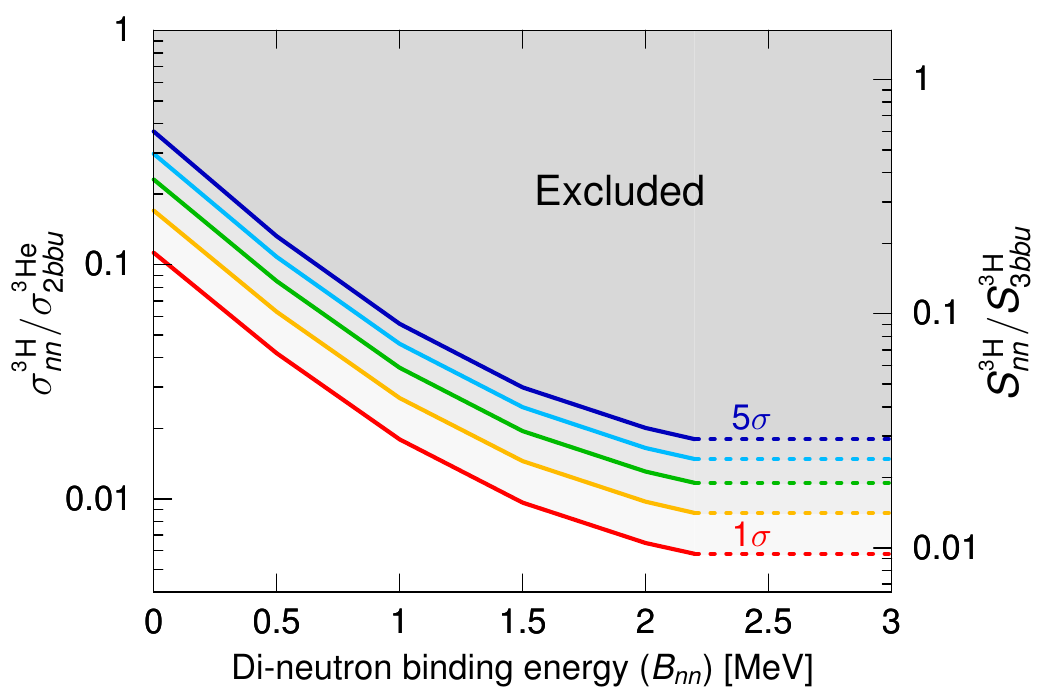}
    \caption{Exclusion limits for a bound dineutron state as a function of binding energy. The left $y$-axis shows the extracted cross section ratio for dineutron production $^3$H$(e,e'p)nn$ relative to 2BBU on $^3$He, i.e., $^3$He$(e,e'p)d$. The right $y$-axis shows the model-dependent estimate for the relative contribution of dineutron break-up in the tritium spectral function.}
    \label{fig:exclusion}
\end{figure}

The results of our exclusion analysis are shown in Fig.~\ref{fig:exclusion}, with the left y-axis showing the ratio of di-neutron cross section from $^3$H relative to the 2bbu cross section from $^3$He in the kinematics measured in this experiment. The ability to exclude di-neutron production depends on di-neutron binding energy. For di-neutrons bound by $>2.2$~MeV, our data exclude a di-neutron cross section of 0.9\% of the $^3$He 2bbu cross section at the $2 \sigma$ confidence level, and exclude a relative di-neutron cross section greater than 2\% at the $5\sigma$ level. Our exclusion limits become less stringent, however, for smaller di-neutron binding energies.

From this cross section ratio we can draw a model-dependent inference about the limits of the relative di-neutron contribution to the $^3$H spectral function. The spectral function, $S(E_{miss},\vec{p}_{miss})$, describes the probability of finding a nucleon in a nucleus with momentum equal to $\vec{p}_{miss}$ and separation energy equal to $E_{miss}$. While our experiment does not have complete acceptance coverage over all $E_{miss}$ and $p_{miss}$, we can compare our measured 2bbu and 3bbu cross sections on $^3$He to a spectral function calculation to derive a correction to account for the incomplete acceptance. Specifically, we aim to place exclusion limits on $S^{^3\text{H}}_{nn} / S^{^3\text{H}}_{3bbu}$, where 
\begin{align}
    S^{^3\text{H}}_{nn} &= \int d^3p_{miss}
    S^{^3\text{H}}(E_{miss}^{nn},\vec{p}_{miss}) \\
        S^{^3\text{H}}_{3bbu} &= \int d^3p_{miss}
    \int_{E_{3bbu}}^\infty d E_{miss} S^{^3\text{H}}(E_{miss},\vec{p}_{miss}).
\end{align}
This ratio can be expanded as
\begin{equation}
    \frac{S^{^3\text{H}}_{nn}}{S^{^3\text{H}}_{3bbu}} =
    \left[\frac{\sigma_{nn}^{^3\text{H}}}{\sigma_{2bbu}^{^3\text{He}}}
    \right]\cdot
        \left (
    \left[\frac{\sigma_{2bbu}^{^3\text{He}}}{\sigma_{3bbu}^{^3\text{He}}}
    \right]\cdot
    \left[
    \frac{{\sigma_{3bbu}^{^3\text{He}}}}{{\sigma_{3bbu}^{^3\text{H}}}}
    \right]\cdot
    \left[
    \frac{S^{^3\text{H}}_{nn}}{S^{^3\text{H}}_{3bbu}}
    \frac{{\sigma_{3bbu}^{^3\text{H}}}}{\sigma_{nn}^{^3\text{H}}}
    \right]
    \right) .
\end{equation}
The first term is the cross section ratio, $R$, on which we placed exclusion limits. The terms enclosed in parentheses represent a correction factor. We can further assume that in the kinematics of the experiment 
\begin{equation*}
     \left[
    \frac{S^{^3\text{H}}_{nn}}{S^{^3\text{H}}_{3bbu}}
    \frac{{\sigma_{3bbu}^{^3\text{H}}}}{\sigma_{nn}^{^3\text{H}}}
    \right] 
    \approx
     \left[
    \frac{S^{^3\text{He}}_{2bbu}}{S^{^3\text{H}e}_{3bbu}}
    \frac{{\sigma_{3bbu}^{^3\text{He}}}}{\sigma_{2bbu}^{^3\text{He}}}
    \right],
\end{equation*}
i.e., that the relative proportions of 2bbu and 3bbu cross sections measured in the experiment versus the underlying spectral function is the same in both nuclei. We can estimate all of the correction factors using a combination of our measured data and a spectral function calculation. In this work, we use the $^3$He spectral function calculation of C.\ Ciofi degli Atti and L.~P.~Kaptari~\cite{CiofidegliAtti:2004jg}. We find that     $(S^{^3\text{He}}_{2bbu}/S^{^3\text{H}e}_{3bbu})/(\sigma_{2bbu}^{^3\text{He}}/\sigma_{3bbu}^{^3\text{He}})=1.24$, and that the entire correction factor is 1.62. We use this factor on the right y-axis of Fig.~\ref{fig:exclusion}, show how our exclusion limits on the cross section translate to limits on the $^3$H spectral function. This estimate suggests that, for a di-neutron binding energy of $\geq 2.2$~MeV, we exclude a $>1.5\%$ dineutron contribution to the $^3$H spectral function at the $2\sigma$ confidence level, and a $>3\%$ contribution at the $5\sigma$ level. Given the heavily model-dependent nature of this approach, we have not quantified its uncertainty. 

\section{Conclusions}

%In agreement with LQCD predictions~\cite{Beane:2011iw} we do not see evidence of a nn-bound state in $^3$H. 
In agreement with our current understanding of the $nn$ interaction, we do not see evidence of a $nn$-bound state in $^3$H.
%%unless the interaction strength is increased and the virtual $nn$ state becomes a bound state~\cite{Deltuva:2019mnv,Beane:2011iw}. 
The discovery of neutral systems as bound or resonant states would have far-reaching implications. Light nuclei that exhibit very asymmetric neutron-to-proton ratios are particularly sensitive to details of the two- and few-body forces used in nuclear models. 

Here, we have used the $^3$He and $^3$H mirror nuclei to look for a possible bound di-neutron system. In the measured kinematics, (high-$Q^2$, $x_B>1$), quasi-elastic electron-induced proton knockout from $^3$He leaves behind a bound deuteron a majority of the time. This clearly identifiable two-body break-up reaction provides a valuable control in the search for a residual di-neutron from $^3$H. Having data from both nuclei allowed us to quantify the measurement's sensitivity without having to rely on modeling the spectrometers' exact performance and resolution.  

In our data, we could not identify a signature for a bound di-neutron with similar binding energy to that of the deuteron.
The experiment's sensitivity allows us to determine that if such a state exists it's appearance is about two orders of magnitude lower than the appearance of the deuteron in the $^3$He case, though this sensitivity degrades rapidly as the di-neutron binding energy decreases.
Other dedicated experiments might be more sensitive in the low energy region, taking advantage of the quasi-elastic scattering demonstrated here, where the distortion of the $nn$ system is minimized, i.e. a recoil-less reaction~\cite{Witala:2010ky}.

\section{Acknowledgements}

This material is based upon work supported by the U.S. Department of Energy, Office of Science, Office of Nuclear Physics under contracts, DE-AC05-06OR23177, DE-SC0016583 and DE-SC0020265.  This work is also supported by the Binational Science Foundation under grant 01025030, the Israel Science Foundation under grant 917/20, the Pazi Foundation, and by NSF grant PHY-1812369



\bibliographystyle{elsarticle-num} 
%\bibliography{inspireshep}
\bibliography{TritiumBib2}

\end{document}